\documentclass[fleqn,usenatbib]{mnras}

\usepackage{newtxtext,newtxmath}

\usepackage[T1]{fontenc}

\DeclareRobustCommand{\VAN}[3]{#2}
\let\VANthebibliography\thebibliography
\def\thebibliography{\DeclareRobustCommand{\VAN}[3]{##3}\VANthebibliography}

\usepackage{graphicx}	
\usepackage{amsmath,bm}	
\usepackage{caption}
\usepackage{subcaption}
\usepackage{orcidlink}

\defcitealias{Dovekie}{Dovekie}
\defcitealias{Grayling24}{G24}
\defcitealias{Thorp21}{T21}

\title[BayeSN $\times$ Dovekie]{BayeSN $\times$ Dovekie: Joint Photometric Cross-calibration and SED Modelling of Type Ia Supernovae}

\author[Grayling et al.]{
M. Grayling$^1$\thanks{Email: mg2102@cam.ac.uk}\orcidlink{0000-0002-6741-983X}, B. Popovic$^2$\orcidlink{0000-0002-8012-6978}, M. Ginolin$^1$\orcidlink{0009-0004-5311-9301}, A. Do$^1$\orcidlink{0000-0003-3429-7845}, K.~S. Mandel$^1$\orcidlink{0000-0001-9846-4417}
\\
$^1$ Institute of Astronomy and Kavli Institute for Cosmology, University of Cambridge, Madingley Road, Cambridge CB3 0HA, UK \\
$^2$ School of Physics and Astronomy, University of Southampton, Southampton, SO17 1BJ, UK \\
}

\date{Accepted XXX. Received YYY; in original form ZZZ}

\pubyear{2026}

\begin{document}
%
%

\def\gPSONELAM{-3.1 \pm 4.2}
\def\gPSONEMAG{\phantom{-}5.8 \pm 3.5}
\def\gPSONEWAV{-0.7}
\def\gPSONEZP{\phantom{-}1.6}
\def\gPSONECORR{\phantom{-}0.024}
\def\rPSONELAM{-5.0 \pm 6.0}
\def\rPSONEMAG{-8.4 \pm 3.0}
\def\rPSONEWAV{-0.8}
\def\rPSONEZP{-2.8}
\def\rPSONECORR{-0.219}
\def\iPSONELAM{-2.1 \pm 5.8}
\def\iPSONEMAG{\phantom{-}0.2 \pm 3.2}
\def\iPSONEWAV{-0.4}
\def\iPSONEZP{\phantom{-}0.1}
\def\iPSONECORR{-0.363}
\def\zPSONELAM{-7.0 \pm 7.8}
\def\zPSONEMAG{\phantom{-}4.4 \pm 3.7}
\def\zPSONEWAV{-0.9}
\def\zPSONEZP{\phantom{-}1.2}
\def\zPSONECORR{-0.243}

\def\gFoundationLAM{-2.8 \pm 3.7}
\def\gFoundationMAG{\phantom{-}3.9 \pm 3.7}
\def\gFoundationWAV{-0.7}
\def\gFoundationZP{\phantom{-}1.0}
\def\gFoundationCORR{-0.037}
\def\rFoundationLAM{\phantom{-}13.4 \pm 5.8}
\def\rFoundationMAG{-5.1 \pm 3.2}
\def\rFoundationWAV{\phantom{-}2.3}
\def\rFoundationZP{-1.6}
\def\rFoundationCORR{-0.206}
\def\iFoundationLAM{-1.6 \pm 6.6}
\def\iFoundationMAG{\phantom{-}0.9 \pm 3.4}
\def\iFoundationWAV{-0.2}
\def\iFoundationZP{\phantom{-}0.3}
\def\iFoundationCORR{-0.283}
\def\zFoundationLAM{-12.2 \pm 14.3}
\def\zFoundationMAG{\phantom{-}1.5 \pm 4.0}
\def\zFoundationWAV{-0.9}
\def\zFoundationZP{\phantom{-}0.4}
\def\zFoundationCORR{-0.009}

\def\gDTHREEYRLAM{\phantom{-}2.4 \pm 4.3}
\def\gDTHREEYRMAG{\phantom{-}2.6 \pm 4.8}
\def\gDTHREEYRWAV{\phantom{-}0.6}
\def\gDTHREEYRZP{\phantom{-}0.5}
\def\gDTHREEYRCORR{\phantom{-}0.016}
\def\rDTHREEYRLAM{\phantom{-}5.0 \pm 4.3}
\def\rDTHREEYRMAG{-6.4 \pm 4.0}
\def\rDTHREEYRWAV{\phantom{-}1.1}
\def\rDTHREEYRZP{-1.6}
\def\rDTHREEYRCORR{-0.129}
\def\iDTHREEYRLAM{\phantom{-}11.5 \pm 7.1}
\def\iDTHREEYRMAG{\phantom{-}1.0 \pm 3.9}
\def\iDTHREEYRWAV{\phantom{-}1.6}
\def\iDTHREEYRZP{\phantom{-}0.3}
\def\iDTHREEYRCORR{-0.265}
\def\zDTHREEYRLAM{-1.1 \pm 11.0}
\def\zDTHREEYRMAG{\phantom{-}4.1 \pm 4.6}
\def\zDTHREEYRWAV{-0.1}
\def\zDTHREEYRZP{\phantom{-}0.9}
\def\zDTHREEYRCORR{-0.367}

\def\CFATHREEKBhLAM{-8.4 \pm 7.4}
\def\CFATHREEKBhMAG{\phantom{-}46.4 \pm 4.6}
\def\CFATHREEKBWAV{-1.1}
\def\CFATHREEKBZP{\phantom{-}10.2}
\def\CFATHREEKBCORR{-0.084}
\def\CFATHREEKVjLAM{\phantom{-}72.7 \pm 5.5}
\def\CFATHREEKVjMAG{\phantom{-}4.1 \pm 3.4}
\def\CFATHREEKVWAV{\phantom{-}13.2}
\def\CFATHREEKVZP{\phantom{-}1.2}
\def\CFATHREEKVCORR{-0.008}
\def\CFATHREEKrkLAM{\phantom{-}53.4 \pm 8.6}
\def\CFATHREEKrkMAG{\phantom{-}7.2 \pm 5.4}
\def\CFATHREEKrWAV{\phantom{-}6.2}
\def\CFATHREEKrZP{\phantom{-}1.3}
\def\CFATHREEKrCORR{-0.715}
\def\CFATHREEKilLAM{-7.5 \pm 7.8}
\def\CFATHREEKilMAG{\phantom{-}26.6 \pm 5.9}
\def\CFATHREEKiWAV{-1.0}
\def\CFATHREEKiZP{\phantom{-}4.5}
\def\CFATHREEKiCORR{-0.531}

\def\CFATHREESBbLAM{\phantom{-}3.2 \pm 13.9}
\def\CFATHREESBbMAG{\phantom{-}5.9 \pm 9.4}
\def\CFATHREESBWAV{\phantom{-}0.2}
\def\CFATHREESBZP{\phantom{-}0.6}
\def\CFATHREESBCORR{-0.041}
\def\CFATHREESVcLAM{\phantom{-}11.4 \pm 10.1}
\def\CFATHREESVcMAG{-9.0 \pm 6.9}
\def\CFATHREESVWAV{\phantom{-}1.1}
\def\CFATHREESVZP{-1.3}
\def\CFATHREESVCORR{\phantom{-}0.002}
\def\CFATHREESRdLAM{\phantom{-}6.9 \pm 13.8}
\def\CFATHREESRdMAG{-2.7 \pm 8.3}
\def\CFATHREESRWAV{\phantom{-}0.5}
\def\CFATHREESRZP{-0.3}
\def\CFATHREESRCORR{-0.424}
\def\CFATHREESIeLAM{-25.5 \pm 16.3}
\def\CFATHREESIeMAG{-0.9 \pm 9.2}
\def\CFATHREESIWAV{-1.6}
\def\CFATHREESIZP{-0.1}
\def\CFATHREESICORR{-0.005}

\def\BCSPLAM{-65.5 \pm 5.7}
\def\BCSPMAG{-2.6 \pm 4.4}
\def\BCSPWAV{-11.6}
\def\BCSPZP{-0.6}
\def\BCSPCORR{-0.145}
\def\VCSPTHREENINELAM{-49.8 \pm 4.9}
\def\VCSPTHREENINEMAG{\phantom{-}1.1 \pm 3.6}
\def\VCSPTHREENINEWAV{-10.1}
\def\VCSPTHREENINEZP{\phantom{-}0.3}
\def\VCSPTHREENINECORR{-0.103}
\def\VCSPTHREEONEFOURLAM{\phantom{-}14.8 \pm 5.3}
\def\VCSPTHREEONEFOURMAG{-31.7 \pm 3.7}
\def\VCSPTHREEONEFOURWAV{\phantom{-}2.8}
\def\VCSPTHREEONEFOURZP{-8.6}
\def\VCSPTHREEONEFOURCORR{\phantom{-}0.046}
\def\VCSPnoshiftLAM{\phantom{-}18.5 \pm 5.4}
\def\VCSPnoshiftMAG{-24.3 \pm 3.7}
\def\VCSPnoshiftWAV{\phantom{-}3.4}
\def\VCSPnoshiftZP{-6.6}
\def\VCSPnoshiftCORR{\phantom{-}0.149}
\def\gCSPLAM{-24.1 \pm 2.8}
\def\gCSPMAG{\phantom{-}5.5 \pm 3.3}
\def\gCSPWAV{-8.7}
\def\gCSPZP{\phantom{-}1.6}
\def\gCSPCORR{-0.062}
\def\rCSPLAM{\phantom{-}24.5 \pm 6.1}
\def\rCSPMAG{-13.6 \pm 4.5}
\def\rCSPWAV{\phantom{-}4.0}
\def\rCSPZP{-3.1}
\def\rCSPCORR{-0.520}
\def\iCSPLAM{-14.1 \pm 8.4}
\def\iCSPMAG{-3.3 \pm 6.6}
\def\iCSPWAV{-1.7}
\def\iCSPZP{-0.5}
\def\iCSPCORR{-0.631}

\def\gSDSSLAM{\phantom{-}3.8 \pm 6.2}
\def\gSDSSMAG{-0.5 \pm 2.5}
\def\gSDSSWAV{\phantom{-}0.6}
\def\gSDSSZP{-0.2}
\def\gSDSSCORR{-0.084}
\def\rSDSSLAM{\phantom{-}16.7 \pm 5.9}
\def\rSDSSMAG{-0.5 \pm 2.8}
\def\rSDSSWAV{\phantom{-}2.8}
\def\rSDSSZP{-0.2}
\def\rSDSSCORR{-0.236}
\def\iSDSSLAM{-11.8 \pm 6.1}
\def\iSDSSMAG{-0.6 \pm 3.1}
\def\iSDSSWAV{-1.9}
\def\iSDSSZP{-0.2}
\def\iSDSSCORR{-0.348}
\def\zSDSSLAM{-28.8 \pm 16.4}
\def\zSDSSMAG{\phantom{-}0.4 \pm 4.2}
\def\zSDSSWAV{-1.8}
\def\zSDSSZP{\phantom{-}0.1}
\def\zSDSSCORR{-0.163}

\def\gSNLSLAM{-7.2 \pm 4.4}
\def\gSNLSMAG{\phantom{-}4.8 \pm 4.8}
\def\gSNLSWAV{-1.6}
\def\gSNLSZP{\phantom{-}1.0}
\def\gSNLSCORR{\phantom{-}0.050}
\def\rSNLSLAM{-41.9 \pm 6.6}
\def\rSNLSMAG{-2.8 \pm 4.0}
\def\rSNLSWAV{-6.3}
\def\rSNLSZP{-0.7}
\def\rSNLSCORR{-0.159}
\def\iSNLSLAM{-8.0 \pm 7.9}
\def\iSNLSMAG{-3.0 \pm 3.7}
\def\iSNLSWAV{-1.0}
\def\iSNLSZP{-0.8}
\def\iSNLSCORR{-0.250}
\def\zSNLSLAM{-15.3 \pm 10.7}
\def\zSNLSMAG{\phantom{-}1.9 \pm 4.6}
\def\zSNLSWAV{-1.4}
\def\zSNLSZP{\phantom{-}0.4}
\def\zSNLSCORR{-0.217}

\def\CFAFOURONEBDLAM{-1.2 \pm 13.7}
\def\CFAFOURONEBDMAG{\phantom{-}3.8 \pm 8.8}
\def\CFAFOURONEBWAV{-0.1}
\def\CFAFOURONEBZP{\phantom{-}0.4}
\def\CFAFOURONEBCORR{-0.059}
\def\CFAFOURONEVELAM{\phantom{-}5.9 \pm 10.0}
\def\CFAFOURONEVEMAG{-3.1 \pm 6.6}
\def\CFAFOURONEVWAV{\phantom{-}0.6}
\def\CFAFOURONEVZP{-0.5}
\def\CFAFOURONEVCORR{-0.116}
\def\CFAFOURONErFLAM{\phantom{-}0.2 \pm 8.3}
\def\CFAFOURONErFMAG{-6.7 \pm 7.3}
\def\CFAFOURONErWAV{\phantom{-}0.0}
\def\CFAFOURONErZP{-0.9}
\def\CFAFOURONErCORR{-0.289}
\def\CFAFOURONEiGLAM{-6.5 \pm 12.5}
\def\CFAFOURONEiGMAG{\phantom{-}0.3 \pm 8.8}
\def\CFAFOURONEiWAV{-0.5}
\def\CFAFOURONEiZP{\phantom{-}0.0}
\def\CFAFOURONEiCORR{-0.311}

\def\CFAFOURTWOBPLAM{-14.0 \pm 16.6}
\def\CFAFOURTWOBPMAG{\phantom{-}3.5 \pm 10.1}
\def\CFAFOURTWOBWAV{-0.8}
\def\CFAFOURTWOBZP{\phantom{-}0.3}
\def\CFAFOURTWOBCORR{-0.031}
\def\CFAFOURTWOVQLAM{-9.1 \pm 13.5}
\def\CFAFOURTWOVQMAG{-6.2 \pm 8.0}
\def\CFAFOURTWOVWAV{-0.7}
\def\CFAFOURTWOVZP{-0.8}
\def\CFAFOURTWOVCORR{-0.048}
\def\CFAFOURTWOrWLAM{-6.8 \pm 13.0}
\def\CFAFOURTWOrWMAG{-7.0 \pm 8.5}
\def\CFAFOURTWOrWAV{-0.5}
\def\CFAFOURTWOrZP{-0.8}
\def\CFAFOURTWOrCORR{-0.289}
\def\CFAFOURTWOiTLAM{\phantom{-}1.9 \pm 13.6}
\def\CFAFOURTWOiTMAG{\phantom{-}1.6 \pm 9.5}
\def\CFAFOURTWOiWAV{\phantom{-}0.1}
\def\CFAFOURTWOiZP{\phantom{-}0.2}
\def\CFAFOURTWOiCORR{-0.093}

\label{firstpage}
\pagerange{\pageref{firstpage}--\pageref{lastpage}}
\maketitle

\begin{abstract}
We present a new framework for BayeSN, the hierarchical Bayesian SED model for type Ia supernovae (SNe Ia), incorporating cross-calibration of samples observed across heterogeneous telescopes. This framework is the first to parametrise the filter wavelength and zero-point offsets commonly used in SN~Ia cosmology within SN SED model training, enabling additional constraint on cross-calibration from SNe beyond the standard stellar-based cross-calibration pipeline. We apply this framework to train a new G26 BayeSN model on the same SED model training sample used in recent cosmological analyses, an order-of-magnitude increase over previous BayeSN training samples, and include a novel training methodology to leverage high-redshift SNe Ia in BayeSN training. We present the G26 model and apply it to the DES-SN5YR sample to assess performance, finding a 12 per cent reduction in $\sigma_{\rm NMAD}$ scatter when compared with SALT3.Dovekie; 0.164 mag compared with 0.185 mag for a sample of likely SNe Ia at $z < 0.7$, without bias corrections. We additionally present constraints on cross-calibration wavelength and zero-point shifts from our framework when using the latest `Dovekie' calibration constraints as a prior. This work is a key step towards a full end-to-end cosmological analysis with BayeSN; the new G26 model is incorporated within the public BayeSN code.
\end{abstract}




\section{Introduction}\label{sec:Intro}


Type Ia supernovae (SNe\,Ia) are valuable cosmological probes, playing a key role in the discovery of the accelerating expansion of the universe \citep{Riess98, Perlmutter99}. In the intervening years, the breadth and depth of the use of SNe\,Ia has only increased, today measuring the dark energy equation-of-state parameter $w$\footnote{One of the potential explanations for this accelerating expansion.} with ever-increasing precision \citep{Betoule14, Brout18SYS, Brout22, DES5YR, Rubin25}. 

A recent combination of SNe\,Ia with complementary probes including Baryon Acoustic Oscillations from \citet{DESI25} has shown potential disagreement with the concordance $\Lambda$CDM cosmological model at significances varying from $2.8$-$4.2\sigma$ \citep[although some studies have suggested that this is a natural consequence of tensions between different cosmological data sets;][]{Ong25}. While still statistically dominated \citep{DES5YR}, measurements of cosmological parameters with SNe\,Ia are increasingly sensitive to systematic uncertainty. \citet{Dhawan24} suggests that this $\Lambda$CDM discrepancy, if not real, could be due to a combination of systematic shifts from the calibration of SNe\,Ia and astrophysical effects such as dust. \citet{Dovekie} (hereafter \citetalias{Dovekie}) has recently attempted to address the calibration side, building on previous cross-calibration attempts from \citet{Fragilistic_pub} and adding a novel calibration pathway using DA WDs from \citet{Boyd25}. In this paper, we build on \citetalias{Dovekie} by incorporating the cross-calibration modelling typically used for SN cosmology directly within the SN Ia SED model BayeSN \citep{Mandel22, Thorp21, Grayling24}; for the first time, we cross-calibrate photometric bands across surveys using SN Ia photometry, using the Dovekie cross-calibration as a prior. Using this treatment, we train a new cross-calibrated optical BayeSN model designed for cosmological inference, which achieves $\sim12$ per cent reduction in scatter compared to a SALT model trained on the same data.

The main source of systematic uncertainty in modern SN cosmology is unknown astrophysics \citep{DES5YR}, in particular the cause of the widely established environmental dependence of SNe Ia \citep[e.g.][]{Sullivan10, Kelly10} and the role that host-galaxy dust extinction or intrinsic differences between SNe in different environments may play in causing this \citep[e.g.][]{Rigault20, BS20, TM22, Popovic22, Wiseman23, Kelsey23, WH23, Grayling24, Ginolin25, Murakami26, Duarte26, Sarin26} \citep[for further discussion please see][]{Grayling24}. One of the main limitations in the method used for standardisation of SNe Ia in recent years, SALT \citep{Guy07, Guy10, Kenworthy21}, is the treatment of colour variation across the population; SALT assumes a single wavelength-dependent function which describes the variation of the underlying SN Ia SED with an apparent colour parameter $c$, roughly corresponding to apparent $B-V$ colour at peak brightness. This approach does not distinguish intrinsic variation across the SN population from the impact of dust extinction, while previous analyses of SN Ia host-galaxy dust properties have shown that this is insufficient to fully model the SN Ia population \citep[e.g.][]{Mandel17, BS20, Popovic22, Ward23, Grayling24, Duarte23, Meldorf23, Hallgren25}. This is one of the key advantages of BayeSN, which includes separate SED-level treatments of dust extinction and intrinsic colour variation.

Despite the benefits that BayeSN presents over SALT, further work is required to prepare BayeSN for a full cosmology analysis. We shall focus on two key tasks here: calibration and training on high-redshift supernovae. Calibration remains the second-largest source of systematic uncertainty \citep{DES5YR, Brout22} in cosmology with SNe Ia today, but BayeSN does not presently take into account the covariance between filter zero points (ZPs) nor the impact of filter wavelength calibration uncertainties. Developing a principled method for including cross-calibration systematics within BayeSN is a crucial step towards cosmological inference. Additionally, BayeSN training traditionally involves assuming a fixed cosmological model; previous models have used low-redshift training samples, meaning assumed distances were only sensitive to the assumed $H_0$, which is not an issue for $w$-cosmology \citep[or for $H_0$ inference, see][]{Dhawan23}. However, training BayeSN on higher-redshift SNe Ia would have become a circular problem; the SED model would already have been conditioned on the cosmological parameters of interest. We present an improved training method which avoids this issue, allowing us to apply BayeSN to a much larger training set.

This work is outlined as follows: in Section \ref{sec:method}, we provide an overview of the BayeSN model and the updates we have made within this work to train our new model. We then present the data samples we use to train and validate this new model in Section \ref{sec:Data}. We present our model and performance on the training sample in Section \ref{sec:Results}, and then consider performance on unseen data from DES-SN5YR in \ref{sec:DES5YR}. Finally, we discuss further steps forward for BayeSN in Section \ref{sec:Conclusions}.

\section{Method}\label{sec:method}

\subsection{Review of the BayeSN Model}

BayeSN is a hierarchical Bayesian model for the time-dependent SEDs of SNe Ia, modelling variations in the intrinsic SED alongside the impact of host-galaxy dust extinction \citep{Mandel22, Thorp21, Grayling24}\footnote{Building upon previous hierarchical multi-band SN~Ia light curve models of \citealt{Mandel_2009,Mandel_2011}.}. A full description of the BayeSN model, including an expression for the full posterior distribution, is derived in \citet{Mandel22}, while further applications of the BayeSN model are presented in e.g. \citet{Thorp21,TM22,Thorp24, W22, Grayling24, Uzsoy24, GP25, Hayes25, BayeSN-TD, Ginolin26}. We present a short review of the BayeSN model here.

The full rest-frame time- and wavelength-dependent BayeSN SED model is given by:
\begin{multline}
\label{BayeSN_eqn}
    -2.5 \log_{10}\left(S_s(t,~\lambda_r)/S_0(t,~\lambda_r)\right) = M_0 +W_0(t,~\lambda_r)  \\ 
    + \delta M_s + \theta_{1,s}W_1(t,~\lambda_r) + \epsilon_s(t,~\lambda_r) + A_{V,s} \, \xi(\lambda_r;~R_{V,s}).
\end{multline}
where $S_0(t,~\lambda_r)$ is the baseline spectral template from \citet{Hsiao11} that spans optical to NIR wavelengths, and $M_0 = -19.5$ is a normalisation constant\footnote{This does not imply a fixed peak absolute magnitude of -$19.5$, as $W_0(t,~\lambda)$ sets the absolute normalisation of a SN Ia SED.}. $W_0(t,~\lambda_r)$ is a warping term that adjusts and normalises the \citet{Hsiao11} template to match the average intrinsic SED of a SN Ia. $W_1(t,~\lambda_r)$ is a functional principal component (FPC), which describes the first mode of intrinsic SED variation across the population; combined with the per-SN parameter $\theta_{1,s}$, these two terms effectively capture the `broader-brighter' relationship established for SNe Ia \citep{Phillips93}. $W_0$, $W_1$ and $\theta_1$ are roughly analogous to $M_0$, $M_1$ and $x_1$ in SALT. Beyond this, $\delta M_s$ is a per-SN `grey'\footnote{Independent of both time and wavelength.} offset, drawn from a Normal distribution $\delta M_s\sim\mathcal{N}(0,~\sigma_0^2)$, where $\sigma_0$ is a hyperparameter inferred during model training which governs the extent of the achromatic intrinsic scatter. $\epsilon_s(t,\lambda_r)$ is a time- and wavelength-dependent function that describes the time-varying residual intrinsic colour variations in the intrinsic SED, represented by spline knots drawn from a multivariate Gaussian $\mathbf{e}_s \sim \mathcal{N}(0, \mathbf{\Sigma}_\epsilon)$, where $\mathbf{e}_s$ is the vectorisation of the $\mathbf{E}_s$ matrix of spline knots. The covariance matrix $\mathbf{\Sigma}_\epsilon$, which describes the population distribution of residual intrinsic scatter, is a hyperparameter of the model which is inferred during training. Finally, $ A_{V,s} \, \xi(\lambda;~R_{V,s})$ is the host-galaxy dust extinction along the line-of-sight for supernova $s$; $A_{V,s}$ is the $V$-band dust extinction that is multiplied by the \citet{Fitzpatrick99} dust law term $\xi(\lambda;~R_{V,s})$. Please note that parameters denoted with $s$ are latent parameters, which differ for every SN, while those without $s$ are hyperparameters shared across the population.

The continuous functions $W_0(t,~\lambda_r)$, $W_1(t,~\lambda_r)$ and $\epsilon_s(t,~\lambda_r)$ are implemented as a cubic spline over two dimensions, described by a matrix of knots; these knots are the model parameters. The spline knots are placed every 10 days from $-10$ through $+40$ days, with rest-frame wavelength knots placed at $(2800,~4900,~6200,~7700,~8700,~9500)$ \AA. These correspond to the central wavelengths of the $griz$ filters as used for the optical BayeSN model presented in \citet{Thorp21}, with the lower knot adjusted bluer to extend the wavelength coverage of the model to match the SALT3.Dovekie model \citepalias{Dovekie}, the model used for the most recent DES-SN5YR cosmological analysis \citep{Popovic26}.

\begin{figure}
    \centering
    \includegraphics[width=9cm]{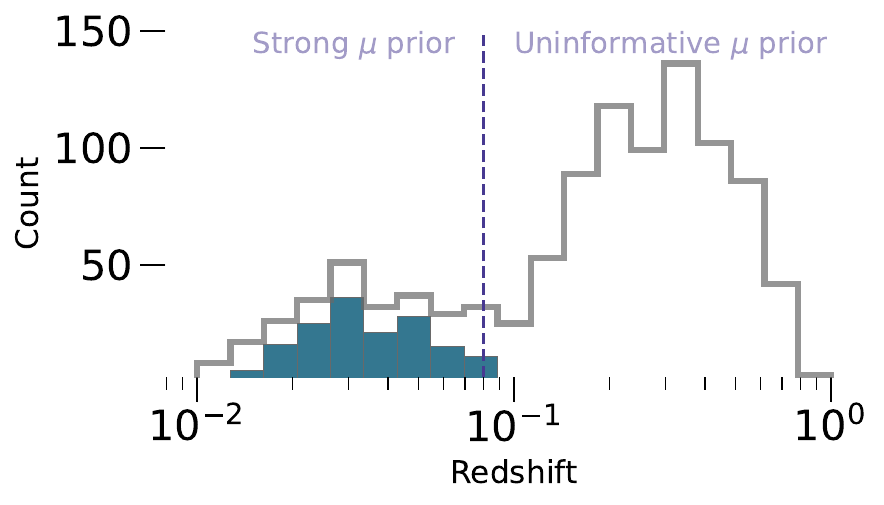}
    \caption{The redshift distribution for the G26 training set (grey) and the \citetalias{Thorp21} training set (blue). We highlight the $z=0.08$ cut where we relax our distance priors (see Section \ref{sec:TrainingMethod}).}
    \label{fig:redshifts}
\end{figure}

\subsection{Calibration Methodology}\label{sec:CalMethod}

\subsubsection{Review of Dovekie}

\citetalias{Dovekie} is a cross-calibration pipeline for combining multiple photometric telescopes across space and time for use in SN Ia cosmology. \citetalias{Dovekie} makes use of tertiary stars provided by completed SN surveys, and uses observations from the Pan-STARRS telescope and Gaia spectra to provide complementary observations. Filter transmission functions are re-determined for each telescope and filter by comparing real observations to standard spectra from CALSPEC and NGSL V2, integrated through provided filter functions from the original surveys. 

After this filter calibration, the stellar photometry is further complemented by the DA WD network from \citet{Boyd25}, and the necessary ZP offsets are calculated across all systems simultaneously using the DA WD and provided stellar photometry. We use the filter response files from \citetalias{Dovekie}, including the wavelength shifts that were applied, within this work.

\subsubsection{Incorporating Cross-Calibration within BayeSN}

Common data releases of SNe Ia light curves \citep{Betoule14,Scolnic18, Sanchez24, Brout22} contain internal calibration, but are not necessarily cross-calibrated with other telescopes\footnote{The JLA data release is a notable exception.}. Large compilations of SNe Ia for use in cosmology, such as Pantheon, Pantheon+ and DES-SN5YR are cross-calibrated, but their data releases do not make \textit{in-situ} changes to the observed light curves based on the cross-calibration. Instead, the necessary changes to ZPs and filter responses are applied during the course of their cosmology pipelines and documented alongside the data. In this work, we properly incorporate these shifts within BayeSN for the first time.

Previous work to include cross-calibration systematics within SN cosmological analyses using SALT has used a Monte Carlo approach; when an SED model is trained, all the filters are perturbed with wavelength and ZP offsets based on results from a stellar cross-calibration framework such as Dovekie. This is repeated multiple times with random realisations of wavelength and ZP offsets for every iteration, creating an ensemble of different SALT models which are each used to fit SN Ia light curves to determine the systematic uncertainty in distances arising from different cross-calibration realisations.

Within this work, we directly include these cross-calibration terms within the BayeSN model. Every filter in the data set the model is applied to has corresponding wavelength shift and ZP offset parameters, which are inferred jointly within the model alongside all other components; this avoids the need for the ensemble approach.

Ignoring cross-calibration terms, a model magnitude from the BayeSN model is given by:

\begin{multline}
    m_{s,i} = -2.5\log_{10}\Big((1+z_s)^{-1}\int f_s(t_i,~\lambda_o)\,\mathcal{B}_i(\lambda_o) \\
    \times\,10^{-0.4\,A_{\mathrm{MW},s}\,\xi(\lambda_o;\,R_{\mathrm{MW}})}\,\lambda_o\,\text{d}\lambda_o\Big)
\end{multline}
where $f_s(t,~\lambda_o)$ is the full BayeSN SED model described in Eq. \ref{BayeSN_eqn}, evaluated in the observer frame and scaled by distance modulus $\mu_s$, and $\mathcal{B}_i(\lambda_o)$ is the filter transmission. $A_{\mathrm{MW},s}\,\xi(\lambda_o;\,R_{\mathrm{MW}})$ is the Milky Way extinction curve for each SN. To incorporate cross-calibration systematics, we add two additional terms when computing model photometry: a wavelength offset in the definition of the filter transmission function and a ZP offset for each filter of each survey analysed. Hereafter, we denote an individual filter of an individual survey with the subscript $T$. In practice, these two offsets modify our model photometry such that:

\begin{multline}
    m_{sT,i} = -2.5\log_{10}\Big((1+z_s)^{-1}\int f_s(t_i,~\lambda_o)\,\mathcal{B}_i(\lambda_o+\Delta\lambda_T) \\
    \times\,10^{-0.4\,A_{\mathrm{MW},s}\,\xi(\lambda_o;\,R_{\mathrm{MW}})}\,\lambda_o\,\text{d}\lambda_o\Big) + \Delta m_T
\end{multline}

Please note that applying a wavelength shift to the filter will in turn modify the integrated effect of Milky Way extinction through the filter; the $\mathcal{B}_i(\lambda_o+\Delta\lambda_T)   \times\,10^{-0.4\,A_{\mathrm{MW},s}\,\xi(\lambda_o;\,R_{\mathrm{MW}})}$ term will change from the nominal case. Previous cross-calibration work from \citet{Fragilistic_pub}\footnote{\citet{Fragilistic_pub} only calculated Milky Way extinction for their field stars, but not for their reference SEDs, leading to inconsistencies in the treatment of Milky Way extinction.} pre-computes Milky Way extinction at the transmission-weighted mean filter wavelength and do not consider the source SED or recompute when shifting the filter. \citetalias{Dovekie}, meanwhile, does not include a treatment of Milky Way extinction\footnote{This was in response to the issue identified in \citet{Fragilistic_pub}, with Milky Way extinction corrections removed after being found to be negligible.}. In this work, we recompute it at every proposed $\Delta\lambda_T$\footnote{In principle, when applying a wavelength shift we should additionally recalculate the filter ZP by integrating the shifted filter through the SED of the reference source for the filter, e.g. an AB or Vega spectrum. However, in this work we include separate $\Delta\lambda_T$ and $\Delta m_T$ terms for consistency with current cosmological analyses.}. This effect is small but not negligible; a 50 \AA\, shift of a $g$-band filter for a typical SN Ia with Milky Way $E(B-V)$ of 0.05 mag causes a change in extinction of $\sim3$ mmag, comparable to the uncertainty on individual ZP shifts in Dovekie. This should be considered when allowing for linear translations of filter responses. When training the BayeSN model presented in this work, we incorporate these additional calibration offsets for each filter present in the training set when evaluating model photometry.

In the equations above, $\Delta m_T$ is the ZP magnitude offset for an individual band. We denote the full vector of magnitude offsets for all bands in the training set as $\mathbf{\Delta m}$. The prior on $\mathbf{\Delta m}$ is taken from the posterior provided by \citetalias{Dovekie}, such that
\begin{equation}
    \mathbf{\Delta m} \sim \mathcal{N}({\bm{\mu}_{\Delta m}},~\bm{\Sigma}_{\Delta m})
\end{equation}
where ${\bm{\mu}_{\Delta m}}$ is the vector of posterior means for the ZP offset for each filter from Dovekie, and $\bm{\Sigma}_{\Delta m}$ is a covariance matrix derived from the Dovekie posterior samples on ZP offsets. The prior on each individual $\Delta \lambda_T$ is also taken from the posterior of Dovekie, such that
\begin{equation}
    \Delta \lambda_T \sim \mathcal{N}({\mu_{\Delta \lambda,T}},~
    \sigma_{\Delta \lambda,T}^2).
\end{equation}
Please note that Dovekie does not incorporate covariance between the wavelength shifts of different bands, or between the wavelength shifts and ZP shifts. We follow this approach in this work, not allowing for these covariances a priori, although as these shifts are jointly inferred within the model there is no such restriction a posteriori.

The methodology we employ here enables the first-ever use of SN data directly to constrain calibration systematics, leveraging the assumption that all SNe Ia across the multiple surveys are drawn from the same underlying distribution\footnote{In principle, this approach could be tested with a variety of transients using more physically-motivated templates, using hierarchical inference enabled by frameworks such as \citet{redback}.}. This incorporation of cross-calibration within the model training represents a significant step towards using BayeSN for end-to-end cosmological inference. 

\subsection{Low-Redshift Training Methodology}\label{sec:TrainingMethod}

Previous BayeSN models have been trained solely on low-redshift SNe Ia, imposing a prior on distance modulus $\mu_s$ based on an assumed cosmological model combined with the redshift uncertainty such that
\begin{equation}
    \mu_s | \, z_s \sim \mathcal{N}(\mu_{\Lambda\text{CDM}}(z_s),~ \sigma^2_{\text{ext},s})
\label{eq:distance_prior}
\end{equation}
where $z_s$ is the spectroscopic redshift of SN $s$, $\mu_{\Lambda\text{CDM}}(z_s)$ is the distance modulus of redshift $z_s$ in our assumed cosmology and $\sigma_{\text{ext},s}$ is the uncertainty in $\mu_{\Lambda\text{CDM}}(z_s)$. This is based on propagating the uncertainty in the redshift $z_s$ through to an uncertainty in $\mu$. The redshift uncertainty is given by the individual measurement uncertainty $\sigma_{z,s}$ added in quadrature with a peculiar velocity dispersion $\sigma_\text{pec}/c$. In this work, and previous BayeSN analyses, we assume $\sigma_{\text{pec}}=150$~km~s$^{-1}$ \citep{Carrick15}. Given that these SNe are all at low redshift, the distance priors are sensitive only to the assumed $H_0$ (with a known simple analytic degeneracy with the absolute magnitude constant) and not other cosmological parameters, e.g. $\Omega_m,~w$. This assumption is a key difference between BayeSN and SALT; when applying a trained BayeSN model to fit SN light curves, BayeSN jointly infers a distance $\mu_s$ simultaneously with other latent SN parameters such as $\theta_{1,s}$, $A_{V,s}$. SALT, meanwhile, is trained independently of cosmology and applies post-hoc relations using summary statistics to estimate distances based on estimates of latent SN parameters ($m_B$, $x_1$, and $c$). 

The approach used by BayeSN has advantages. Distances are inferred using the full SN light curve across all bands, rather than based on compressed summary statistics. Inferred distances are also marginalised over the population distribution of dust and intrinsic colour. However, this approach also poses challenges when training the BayeSN model on higher-redshift SNe where distances are sensitive to cosmological parameters other than $H_0$. If we take the same approach, the problem of cosmological parameter inference becomes circular; the model has already been conditioned on cosmological distances at redshifts sensitive to those parameters.

One option would be to reparametrise BayeSN to adopt a SALT-like approach, using latent SN parameters in a post-hoc relation to estimate distance and avoiding imposing a distance prior during training. However, this would remove some of the key strengths of BayeSN as a method of inferring distance. Instead, in this work we adopt a hybrid approach whereby we impose the informative distance prior from Eq. \ref{eq:distance_prior} \textit{only} for SNe at redshifts low enough that the cosmological distance is not sensitive to cosmological parameters other than $H_0$; in this work we set the threshold for this at $z<0.08$. For SNe above this, we instead use an uninformative flat prior. In this way, we maintain BayeSN in its current form while avoiding the circularity of conditioning the model on cosmological parameters.

\section{Data}\label{sec:Data}

We now introduce the data to which we apply BayeSN in this work. We use the collection of SNe from \citet{Kenworthy21} with selection and calibration updates from \citetalias{Dovekie} to train a new cross-calibrated optical BayeSN model, hereafter called the G26 model. In short, we combine data from Foundation, CSP, CfA3+4, DES, SDSS, PS1, and SNLS. Here, we provide a general overview of the selected SNe and the respective surveys; the total number of SNe and spectra for each survey, alongside filters and sources, is given in Table \ref{tab:training_set}. Fig. \ref{fig:redshifts} shows the redshift distribution of our training set, compared with the previous optical-only BayeSN model from \citet{Thorp21}.

\subsection{Low-Redshift Surveys}\label{sec:Data:subsec:low}

At low redshift, we include a coterie of nearby surveys: the Carnegie Supernova Project (CSP, \citealp{krisciunis17}), Foundation \citep{Foley18}, and Center for Astrophysics (CfA, \citealp{Hicken09a, Hicken09b, Hicken12}). These nearby surveys provide a strong base of photometric and spectroscopic data points. CSP, while only containing 13 unique SNe, provides 36 spectra across a wide wavelength range extending into the near infrared. Foundation provides good photometric coverage of the optical $griz$ bands alongside spectroscopic data, while CfA3 and 4 provide valuable spectral time series of SNe Ia. 

\subsection{PS1, SDSS, SNLS, DES}\label{sec:Data:subsec:high}

The Dark Energy Survey (DES, \citealp{Abbott19}), Sloan Digital Sky Survey (SDSS, \citealp{Holtzman08}), Pan-STARRS medium deep (PS1, \citealp{Jones17}), and SuperNova Legacy Survey (SNLS, \citealp{Astier06}) comprise the `high redshift' ($z > 0.1$) surveys in the training sample. SNLS provides an excellent sample of spectra at high redshift $z\sim 0.85$, and alongside DES, probes rest-frame near-UV wavelengths, complementing the mid-$z$ coverage from SDSS and PS1.

\begin{table}
    \centering
    \begin{tabular}{l|ccc}
        Survey & $N_{\rm SNe}$ & Filters & Reference \\
        \hline
        Foundation & 153 & $griz$ & \citet{Foley18} \\
        CfA3 & 44 & $BVRIri$ & \citet{Hicken09a} \\
        CfA4 & 30 & $BVri$ & \citet{Hicken12} \\
        CSP & 12 & $BVgri$ & \citet{krisciunis17} \\
        SDSS & 202  & $griz$ & \citet{Holtzman08} \\
        PS1 & 266 & $griz$ & \citet{Scolnic18} \\
        DES & 206 & $griz$ & \citet{Abbott19} \\
        SNLS & 111 & $griz$ & \citet{Astier06} \\
        \hline
        \textbf{Total} & 1024 & - & -
    \end{tabular}
    \caption{Summary of the G26 model training sample.}
    \label{tab:training_set}
\end{table}

\subsection{DES-SN5YR}

We additionally apply our trained BayeSN model to the DES-SN5YR sample, to provide a validation sample independent of the model training data. The full release of the final DES SN program light curves from \citet{Sanchez24} encompassed the 5 years of observation with the Dark Energy Camera \citep{Flaugher15}. These observations, across ten 3 deg$^2$ fields, went to a depth of $\sim$23.5 mag in $griz$ bands ($\sim$24.5 for the three deep fields). Further details are given in \citet{Kessler15} and \citet{Smith20}.

\section{Model Training Results}\label{sec:Results}

We now present results from the new G26 BayeSN model, trained on the 1024 SNe Ia presented in Table \ref{tab:training_set}, incorporating our new cross-calibration method with BayeSN. The model is trained using the publicly available GPU-accelerated BayeSN code outlined in \citet{Grayling24}. We assess the performance of this new trained model, in terms of Hubble residual (HR) scatter, and present the posterior distributions on cross-calibration shifts derived from our BayeSN model. To assess the performance of our new trained BayeSN model, we compare the distances we infer with this model to SALT distances from \citetalias{Dovekie}.

\subsection{The G26 Model}

\begin{figure}
    \centering
    \includegraphics[width = \columnwidth]{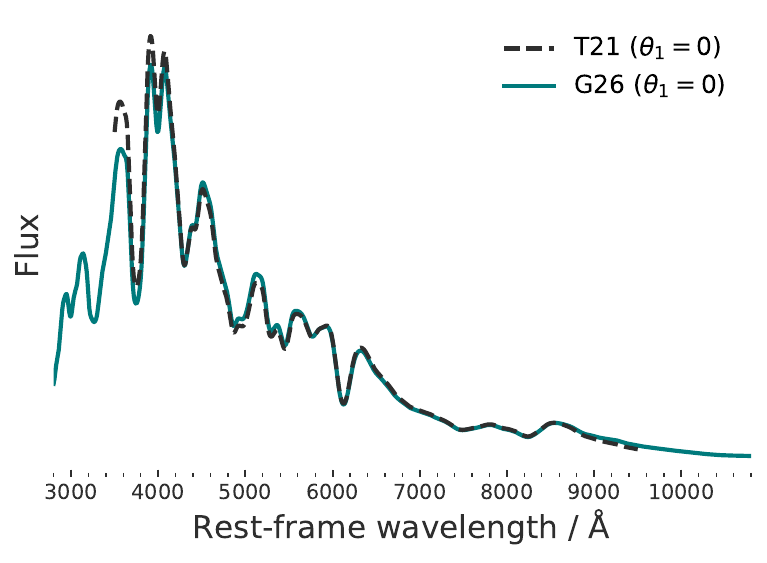}
    \caption{Model SED at peak for the G26 BayeSN model, plotted alongside that of the T21 model to demonstrate the enhanced coverage of blue wavelengths for this optical BayeSN model.}
    \label{fig:G26vsT21}
\end{figure}

\begin{figure}
    \centering
    \includegraphics[width = \columnwidth]{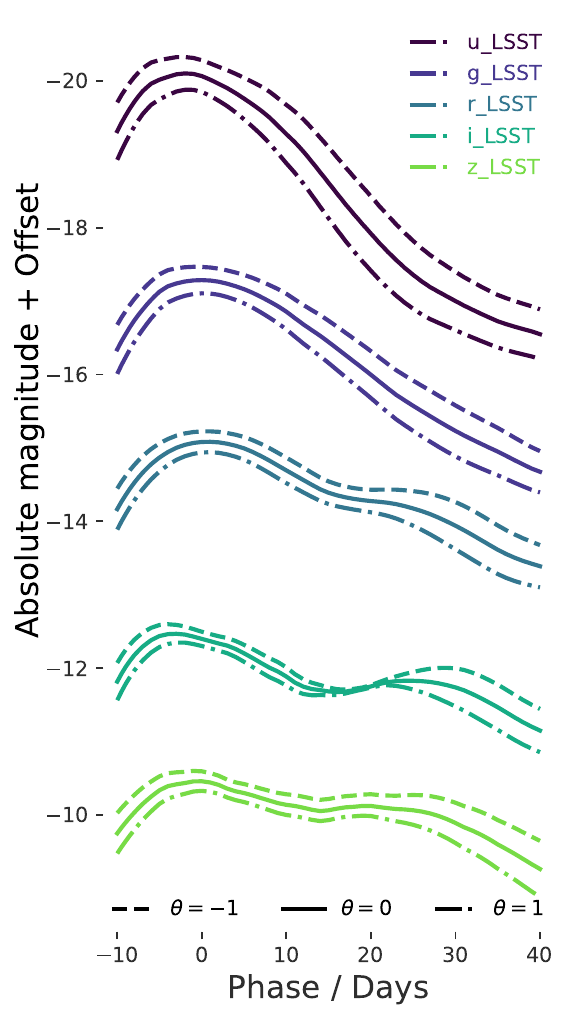}
    \caption{Model rest-frame light curves for the G26 BayeSN model for $ugriz$ photometric bands for a range of $\theta_1$ values.}
    \label{fig:G26lcs}
\end{figure}

The SED surface at peak for a $\theta_1=0$ SN is presented in Fig. \ref{fig:G26vsT21}, shown alongside the same for the T21 optical BayeSN model \citep{Thorp21}. Most apparent is that the G26 model has greater wavelength coverage, particularly at bluer wavelengths; the blue edge of the model is 2800 \AA\ compared to 3500 \AA\ for T21. Note that the model is not trained on any $u/U$-band data, as these are generally considered unreliable for precision cosmology and not included in Dovekie. Nevertheless, extending the training set to higher redshifts means that other bands can cover rest-frame $u$-band which has allowed us to improve the wavelength coverage of the model. Fig. \ref{fig:G26lcs} shows rest-frame model $ugriz$ light curves, demonstrating how the model looks photometrically and how it varies with the light curve shape parameter $\theta_1$. The inclusion of $u$-band here further demonstrates the benefit of the wider wavelength coverage.

There have been a variety of BayeSN models presented in recent years. The M20 model \citep{Mandel22} was an optical-NIR model trained on 79 SNe from CfA and CSP, while the T21 model \citep{Thorp21} was an optical-only model trained on 157 SNe from Foundation. The W22 model \citep{W22} was an optical-NIR model trained on the combination of these two data sets. Most recently, \citet{BayeSN-TD} presented a phase-extended optical-NIR BayeSN model (G25-extended) with coverage out to +85 days, motivated by the need to fit late-time observations of strongly lensed SNe Ia; unlike others, this model also included $U$-band data in training to ensure full wavelength coverage for analysis of SN H0pe \citep{Frye23, Pierel24}. In this context, the G26 model is designed to be a cosmology-grade optical model for precision analyses, while the G25-extended and W22 models represent the current state-of-the-art of optical-NIR modelling of SNe Ia.

Previous publicly-released BayeSN models have used a single value of host galaxy $R_V$ across all SNe \citep[e.g.][]{Mandel22, Thorp21, W22}, an approach more similar to the single colour law used by SALT\footnote{Although SALT has a single colour law to model both intrinsic variation and dust extinction, unlike BayeSN which models each effect separately.}. To be clear, this refers to available pre-trained models; numerous previous BayeSN works \citep[e.g.][]{Thorp21, TM22, Grayling24, Ginolin26} have implemented extended models to study population distributions of $R_{V,s}$. For the G26 model, we instead opt for a Gaussian $R_V$ population distribution as part of the model; fitting distances using this model involves marginalising over a prior distribution of $R_V$. Although observations of most SNe Ia do not have sufficient wavelength coverage to obtain strong constraints on $R_V$, this approach allows for $R_{V,s}$ values away from the population prior where supported by data. Several previous BayeSN analyses \citep{Grayling24, GP25, Ginolin26} have analysed volume-limited samples to constrain properties of the full population of SNe Ia. However, in this work we train the G26 model on the full training set used for SALT3.Dovekie covering a variety of SN surveys, each with their own selection effects. Thus, the dust properties inferred are the effective properties for this sample and should not be regarded as representative of the full population of SNe Ia. Our objective in this work is to train an SED model on the largest sample to minimise scatter in the population, just as the SALT training process does not incorporate a treatment of selection effects. For cosmological inference, this model will be combined with more representative BayeSN forward simulations to calculate bias corrections \citep{Kessler16} as used in recent analyses such as \citet{Brout22, Popovic26}. In future, we will work towards incorporating selection effects in the BayeSN model using simulation-based inference (SBI) \citep[e.g.][]{Karchev24, Boyd26, Karchev26} for joint inference of cosmological parameters and properties of the population of SNe Ia. For this sample, we infer $\mu_R=1.84\pm0.20$, 68th (95th) percentile upper limits of $\sigma_R < 0.28 (0.58)$ and $\tau_A=0.17\pm0.01$ mag, though these should be interpreted considering the caveats above.

\subsection{Hubble Diagram Scatter}

We begin by applying our trained BayeSN model to fit light curves from the surveys with SNe included in the training set. In total we consider fits to 1157 SNe, covering most of the SNe included in SALT3.Dovekie training but extending the sample slightly further; we choose this sample to compare with a readily available set of SALT fits. We measure the inferred scatter in the Hubble diagram using the Normalised Median Absolute Deviation ($\sigma_{\rm NMAD}$),
\begin{equation}
    \sigma_{\rm NMAD} = 1.48 \times (\rm median(|\mu - \mu_{\rm model}|)).
\end{equation}
Table \ref{tab:scatter_survey} presents the Hubble diagram scatter from these fits for each survey within the training set, from both our new BayeSN model and the SALT.Dovekie model. When applied to previously seen SNe, BayeSN has a consistent scatter of 0.135 mag compared to 0.136 mag for SALT3.Dovekie. This demonstrates that our model is able to match the current standard for cosmological inference across a wide range of surveys and redshifts, although results for the G26 model are more impressive when applied to completely unseen data as discussed in Section \ref{sec:DES5YR}.

\begin{table}
    \centering
    \begin{tabular}{l|ccc}
          Survey & median $z$ & $\sigma_{\rm NMAD}$ (G26) & $\sigma_{\rm NMAD}$ (SALT) \\
          & & / mag & / mag \\
          \hline
          Foundation & $0.03$ & 0.134 & 0.136 \\
          CfA3       & $0.03$ & 0.140 & 0.120 \\
          CfA4       & $0.03$ & 0.114 & 0.087 \\
          CSP        & $0.03$ & 0.106 & 0.147 \\
          SDSS       & $0.19$ & 0.132 & 0.130 \\
          PS1        & $0.29$ & 0.140 & 0.145 \\
          DES        & $0.33$ & 0.142 & 0.135 \\
          SNLS       & $0.52$ & 0.129 & 0.139 \\
          \hline
          Total      & --     & 0.135 & 0.136 \\
    \end{tabular}   
    \caption{The Hubble residual scatter (NMAD) for each survey when fit with the G26 and SALT3.Dovekie models. We provide the median redshift of each survey.}
    \label{tab:scatter_survey}
\end{table}

\subsection{Calibration with BayeSN}

We now present our inferred constraints on cross-calibration, using the result from Dovekie as a prior within our BayeSN cross-calibration framework. To test the robustness of our cross-calibration approach, we include a validation test where we inject artificial shifts in pre-processing to well-constrained filters and verify that the code is able to undo these artificial shifts and recover the original solution. This analysis is presented in detail in Appendix \ref{appendix:validation}; in brief, the code is successfully able to reverse our artificial shifts and recover the expected solutions.

\begin{figure}
    \centering
    \includegraphics[width = \columnwidth]{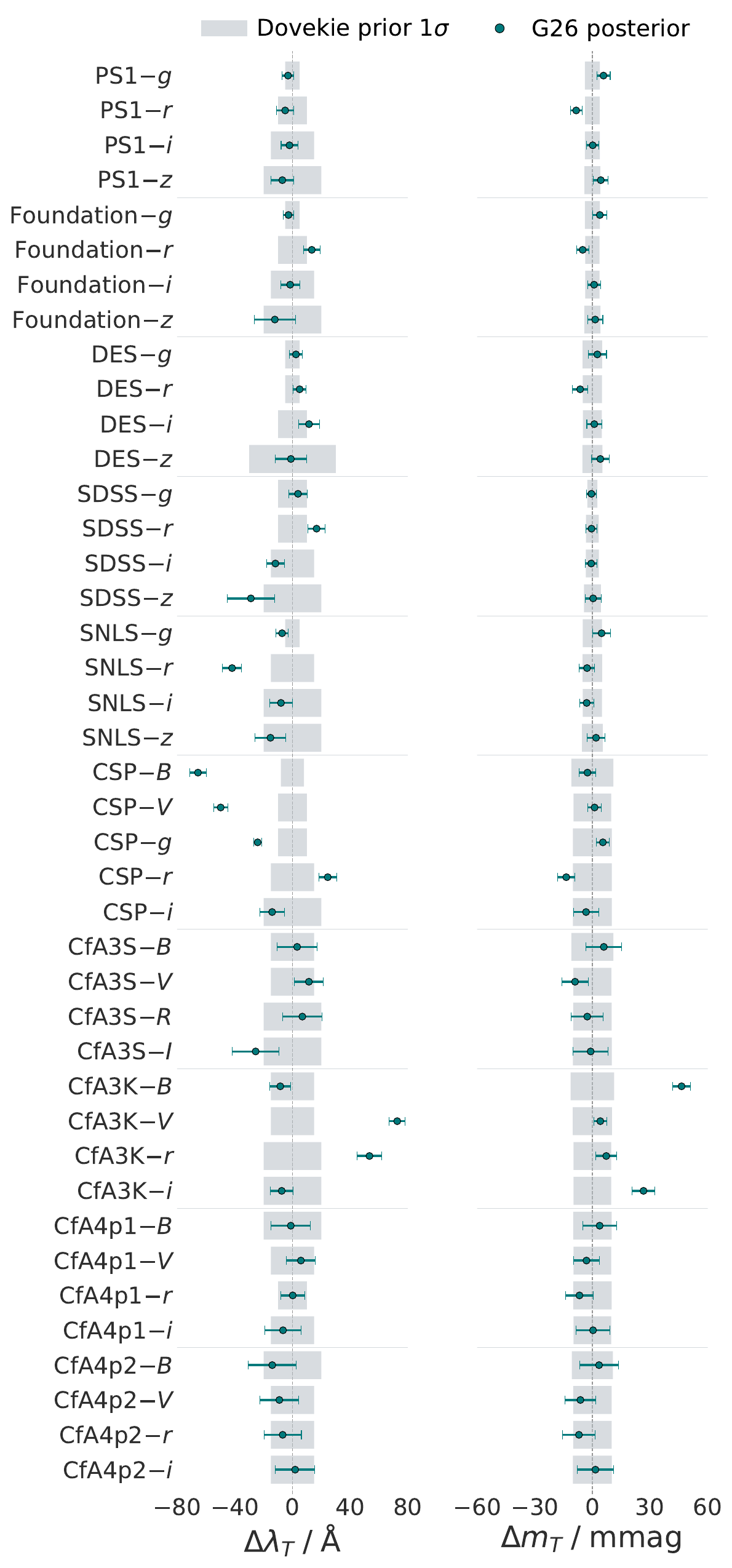}
    \caption{Posterior distributions on our inferred wavelength and ZP shifts using BayeSN, compared with prior distributions from Dovekie. The posteriors are represented by the mean and standard deviation, while the shaded regions represent the standard deviation of the prior from Dovekie. In the case of the multivariate prior used for the ZP shifts, these shaded regions represent the square root of the diagonal elements of the covariance matrix.}
    \label{fig:calibration_shifts}
\end{figure}

Fig. \ref{fig:calibration_shifts} presents posteriors for $\mathbf{\Delta\lambda}$ and $\mathbf{\Delta m}$ from our BayeSN-based calibration method, alongside priors from \citetalias{Dovekie}, following the methods described in Section \ref{sec:CalMethod}; these are also detailed in Table \ref{tab:calibrationresults}. For the most part, we find $\mathbf{\Delta\lambda}$ posteriors consistent with our priors: PS1, Foundation, DES, SNLS, and CfA4 agree within $\sim1-2\sigma$ from our $\mathbf{\Delta\lambda}$ priors. We do find a few notable outliers: CfA3K$-V$, CfA3K$-r$, CSP$-B$, CSP$-g$, and SNLS$-r$, with wavelength shifts at a significance of $>5\sigma$. Of note, only one of these shifts `reverts' the original filter shift made by Dovekie: the CSP$-B$ band is shifted here back to the original CSP data release value. In many cases our posteriors on $\mathbf{\Delta m}$ are significantly tighter than our priors, most notably in redder filters which tend to have higher prior uncertainty. This suggests that SN data is providing additional constraint on cross-calibration beyond the Dovekie prior, rather than the prior constraints simply being marginalised over when training the model.

\begin{table*}
    \centering
    \caption{Marginal priors and posteriors on filter wavelength and ZP offsets, along with the Pearson correlation coefficient $r_{\lambda,m}$ between posterior samples of $\Delta\lambda_T$ and $\Delta m_T$ to describe the strength of the covariance between the two. All priors have a mean of zero, as they are taken from the results of the Dovekie cross-calibration.}
    \begin{tabular}{l|cccc|c}
         Filter & \multicolumn{2}{c|}{$\Delta\lambda_T$ / \AA} & \multicolumn{2}{c|}{$\Delta m_T$ / mmag} & $r_{\lambda,m}$ \\
          & Prior & Posterior & Prior & Posterior & \\
         \hline
         PS1$-g$ & $\mathcal{N}(0, 5^2)$ & $\gPSONELAM$  & $\mathcal{N}(0, 3.9^2)$ & $\gPSONEMAG$   & ${\gPSONECORR}$ \\
         PS1$-r$ & $\mathcal{N}(0, 10^2)$ & $\rPSONELAM$ & $\mathcal{N}(0, 3.8^2)$ & $\rPSONEMAG$  & ${\rPSONECORR}$ \\
         PS1$-i$ & $\mathcal{N}(0, 15^2)$ & $\iPSONELAM$ & $\mathcal{N}(0, 3.9^2)$ & $\iPSONEMAG$   & ${\iPSONECORR}$ \\
         PS1$-z$ & $\mathcal{N}(0, 20^2)$ & $\zPSONELAM$  & $\mathcal{N}(0, 4.1^2)$ & $\zPSONEMAG$   & ${\zPSONECORR}$ \\
         Foundation$-g$ & $\mathcal{N}(0, 5^2)$ & $\gFoundationLAM$ & $\mathcal{N}(0, 3.9^2)$ & $\gFoundationMAG$  & ${\gFoundationCORR}$ \\
         Foundation$-r$ & $\mathcal{N}(0, 10^2)$ & $\rFoundationLAM$ & $\mathcal{N}(0, 3.7^2)$ & $\rFoundationMAG$ & ${\rFoundationCORR}$ \\
         Foundation$-i$ & $\mathcal{N}(0, 15^2)$ & $\iFoundationLAM$ & $\mathcal{N}(0, 3.7^2)$ & $\iFoundationMAG$  & ${\iFoundationCORR}$ \\
         Foundation$-z$ & $\mathcal{N}(0, 20^2)$ & $\zFoundationLAM$ & $\mathcal{N}(0, 4.2^2)$ & $\zFoundationMAG$  & ${\zFoundationCORR}$ \\
         \textcolor{white}{s} & \textcolor{white}{s} & \textcolor{white}{s} & \textcolor{white}{s} & \textcolor{white}{s} & \textcolor{white}{s} \\
         DES$-g$ & $\mathcal{N}(0, 5^2)$ & $\gDTHREEYRLAM$ & $\mathcal{N}(0, 5.1^2)$ & $\gDTHREEYRMAG$  & ${\gDTHREEYRCORR}$ \\
         DES$-r$ & $\mathcal{N}(0, 5^2)$ & $\rDTHREEYRLAM$ & $\mathcal{N}(0, 5.0^2)$ & $\rDTHREEYRMAG$   & ${\rDTHREEYRCORR}$ \\
         DES$-i$ & $\mathcal{N}(0, 10^2)$ & $\iDTHREEYRLAM$ & $\mathcal{N}(0, 4.9^2)$ & $\iDTHREEYRMAG$  & ${\iDTHREEYRCORR}$ \\
         DES$-z$ & $\mathcal{N}(0, 30^2)$ & $\zDTHREEYRLAM$ & $\mathcal{N}(0, 5.2^2)$ & $\zDTHREEYRMAG$   & ${\zDTHREEYRCORR}$ \\
         \textcolor{white}{s} & \textcolor{white}{s} & \textcolor{white}{s} & \textcolor{white}{s} & \textcolor{white}{s} & \textcolor{white}{s} \\
         CfA3K$-B$ & $\mathcal{N}(0, 15^2)$ & $\CFATHREEKBhLAM$ & $\mathcal{N}(0, 11.3^2)$ &  $\CFATHREEKBhMAG$  & ${\CFATHREEKBCORR}$ \\ 
         CfA3K$-V$ & $\mathcal{N}(0, 15^2)$ & $\CFATHREEKVjLAM$ & $\mathcal{N}(0, 10.2^2)$ & $\CFATHREEKVjMAG$  & ${\CFATHREEKVCORR}$ \\ 
         CfA3K$-r$ & $\mathcal{N}(0, 20^2)$ & $\CFATHREEKrkLAM$ & $\mathcal{N}(0, 9.8^2)$ & $\CFATHREEKrkMAG$  & ${\CFATHREEKrCORR}$ \\ 
         CfA3K$-i$ & $\mathcal{N}(0, 20^2)$ & $\CFATHREEKilLAM$ & $\mathcal{N}(0, 9.8^2)$ & $\CFATHREEKilMAG$  & ${\CFATHREEKiCORR}$ \\ 
         \textcolor{white}{s} & \textcolor{white}{s} & \textcolor{white}{s} & \textcolor{white}{s} & \textcolor{white}{s} & \textcolor{white}{s} \\
         CfA3S$-B$ & $\mathcal{N}(0, 15^2)$ & $\CFATHREESBbLAM$  & $\mathcal{N}(0, 10.9^2)$ & $\CFATHREESBbMAG$  & ${\CFATHREESBCORR}$ \\
         CfA3S$-V$ & $\mathcal{N}(0, 15^2)$ & $\CFATHREESVcLAM$  & $\mathcal{N}(0, 9.9^2)$ & $\CFATHREESVcMAG$  & ${\CFATHREESVCORR}$ \\
         CfA3S$-R$ & $\mathcal{N}(0, 20^2)$ &  $\CFATHREESRdLAM$ & $\mathcal{N}(0, 10.0^2)$ & $\CFATHREESRdMAG$   & ${\CFATHREESRCORR}$ \\
         CfA3S$-I$ & $\mathcal{N}(0, 20^2)$ & $\CFATHREESIeLAM$ & $\mathcal{N}(0, 10.1^2)$ & $\CFATHREESIeMAG$  & ${\CFATHREESICORR}$ \\ %
         \textcolor{white}{s} & \textcolor{white}{s} & \textcolor{white}{s} & \textcolor{white}{s} & \textcolor{white}{s} & \textcolor{white}{s} \\
         CSP$-B$ & $\mathcal{N}(0, 8^2)$ & $\BCSPLAM$ & $\mathcal{N}(0, 10.9^2)$ & $\BCSPMAG$  & ${\BCSPCORR}$ \\
         CSP$-V$ & $\mathcal{N}(0, 10^2)$ & $\VCSPTHREENINELAM$ & $\mathcal{N}(0, 9.8^2)$ & $\VCSPTHREENINEMAG$  & ${\VCSPTHREENINECORR}$ \\
         CSP$-g$ & $\mathcal{N}(0, 10^2)$ & $\gCSPLAM$ & $\mathcal{N}(0, 10.1^2)$ & $\gCSPMAG$  & ${\gCSPCORR}$ \\
         CSP$-r$ & $\mathcal{N}(0, 15^2)$ & $\rCSPLAM$ & $\mathcal{N}(0, 10.1^2)$ & $\rCSPMAG$  & ${\rCSPCORR}$ \\
         CSP$-i$ & $\mathcal{N}(0, 20^2)$ & $\iCSPLAM$ & $\mathcal{N}(0, 10.1^2)$ & $\iCSPMAG$  & ${\iCSPCORR}$ \\
         \textcolor{white}{s} & \textcolor{white}{s} & \textcolor{white}{s} & \textcolor{white}{s} & \textcolor{white}{s} & \textcolor{white}{s} \\
         SDSS$-g$ & $\mathcal{N}(0, 10^2)$ & $\gSDSSLAM$ & $\mathcal{N}(0, 2.6^2)$ & $\gSDSSMAG$  & ${\gSDSSCORR}$ \\
         SDSS$-r$ & $\mathcal{N}(0, 10^2)$ & $\rSDSSLAM$ & $\mathcal{N}(0, 3.3^2)$ & $\rSDSSMAG$  & ${\rSDSSCORR}$ \\
         SDSS$-i$ & $\mathcal{N}(0, 15^2)$ & $\iSDSSLAM$ & $\mathcal{N}(0, 3.4^2)$ & $\iSDSSMAG$  & ${\iSDSSCORR}$ \\
         SDSS$-z$ & $\mathcal{N}(0, 20^2)$ & $\zSDSSLAM$ & $\mathcal{N}(0, 4.4^2)$ & $\zSDSSMAG$  & ${\zSDSSCORR}$ \\
         \textcolor{white}{s} & \textcolor{white}{s} & \textcolor{white}{s} & \textcolor{white}{s} & \textcolor{white}{s} & \textcolor{white}{s} \\
         SNLS$-g$ & $\mathcal{N}(0, 5^2)$ & $\gSNLSLAM$ & $\mathcal{N}(0, 5.0^2)$ & $\gSNLSMAG$  & ${\gSNLSCORR}$ \\
         SNLS$-r$ & $\mathcal{N}(0, 15^2)$ & $\rSNLSLAM$ & $\mathcal{N}(0, 5.1^2)$ & $\rSNLSMAG$  & ${\rSNLSCORR}$ \\
         SNLS$-i$ & $\mathcal{N}(0, 20^2)$ & $\iSNLSLAM$ & $\mathcal{N}(0, 5.0^2)$ & $\iSNLSMAG$  & ${\iSNLSCORR}$ \\
         SNLS$-z$ & $\mathcal{N}(0, 20^2)$ & $\zSNLSLAM$ & $\mathcal{N}(0, 5.4^2)$ & $\zSNLSMAG$  & ${\zSNLSCORR}$ \\
         \textcolor{white}{s} & \textcolor{white}{s} & \textcolor{white}{s} & \textcolor{white}{s} & \textcolor{white}{s} & \textcolor{white}{s} \\
         CfA4p1$-B$ & $\mathcal{N}(0, 20^2)$ & $\CFAFOURONEBDLAM$ & $\mathcal{N}(0, 10.0^2)$ & $\CFAFOURONEBDMAG$  & ${\CFAFOURONEBCORR}$ \\ 
         CfA4p1$-V$ & $\mathcal{N}(0, 15^2)$ & $\CFAFOURONEVELAM$ & $\mathcal{N}(0, 9.8^2)$ & $\CFAFOURONEVEMAG$  & ${\CFAFOURONEVCORR}$ \\ 
         CfA4p1$-r$ & $\mathcal{N}(0, 10^2)$ & $\CFAFOURONErFLAM$ & $\mathcal{N}(0, 10.0^2)$ & $\CFAFOURONErFMAG$  & ${\CFAFOURONErCORR}$ \\ 
         CfA4p1$-i$ & $\mathcal{N}(0, 15^2)$ & $\CFAFOURONEiGLAM$ & $\mathcal{N}(0, 9.8^2)$ & $\CFAFOURONEiGMAG$  & ${\CFAFOURONEiCORR}$ \\ 
         \textcolor{white}{s} & \textcolor{white}{s} & \textcolor{white}{s} & \textcolor{white}{s} & \textcolor{white}{s} & \textcolor{white}{s} \\
         CfA4p2$-B$ & $\mathcal{N}(0, 20^2)$ & $\CFAFOURTWOBPLAM$ & $\mathcal{N}(0, 10.7^2)$ & $\CFAFOURTWOBPMAG$  & ${\CFAFOURTWOBCORR}$ \\ 
         CfA4p2$-V$ & $\mathcal{N}(0, 15^2)$ & $\CFAFOURTWOVQLAM$ & $\mathcal{N}(0, 10.0^2)$ & $\CFAFOURTWOVQMAG$  & ${\CFAFOURTWOVCORR}$ \\ 
         CfA4p2$-r$ & $\mathcal{N}(0, 15^2)$ & $\CFAFOURTWOrWLAM$ & $\mathcal{N}(0, 10.0^2)$ & $\CFAFOURTWOrWMAG$  & ${\CFAFOURTWOrCORR}$ \\ 
         CfA4p2$-i$ & $\mathcal{N}(0, 15^2)$ & $\CFAFOURTWOiTLAM$ & $\mathcal{N}(0, 10.1^2)$ & $\CFAFOURTWOiTMAG$  & ${\CFAFOURTWOiCORR}$ \\ 
         \textcolor{white}{s} & \textcolor{white}{s} & \textcolor{white}{s} & \textcolor{white}{s} & \textcolor{white}{s} & \textcolor{white}{s} \\
    \end{tabular}
    \label{tab:calibrationresults}
\end{table*}

Our results are similar when considering $\mathbf{\Delta m}$. Overall our results are typically consistent with the Dovekie prior, although with a few notable outliers in CfA3K $BVi$ bands. The posterior widths are also generally similar to the prior, with a slight reduction of about $1-2$ mmag across most of the sample, although for CfA3K and CfA in particular there is a notable increase in precision of $\sim5$ mmag across all filters when including the SN data, compared with the Dovekie prior.

Table \ref{tab:calibrationresults} additionally presents the Pearson correlation coefficient $r_{\lambda,m}$ as a measure of the strength of the posterior covariance between $\mathbf{\Delta\lambda}$ and $\mathbf{\Delta m}$ for each band. The approach used by \citetalias{Dovekie} infers the wavelength and ZP offsets in a two-stage method and does not consider covariance between the two. We consider the posterior correlation in this way to assess the validity of this two-stage approach. Overall, this approximation does seem reasonable in most cases, with 56 per cent of filters having $|r|<0.2$. There are a small number of filters where these are highly covariant, most notably CfA3K-r and CSP-i bands. For $i$-band, this follows naturally from the presence of telluric lines in the atmospheric transmission. Ideally this would not shift alongside the rest of the filter transmission function, although in practice it does as current approaches do not model the atmospheric term separately from the underlying filter transmission. 

\section{Independent Validation with DES-SN5YR}
\label{sec:DES5YR}

\begin{figure}
    \centering
    \includegraphics[width = \columnwidth]{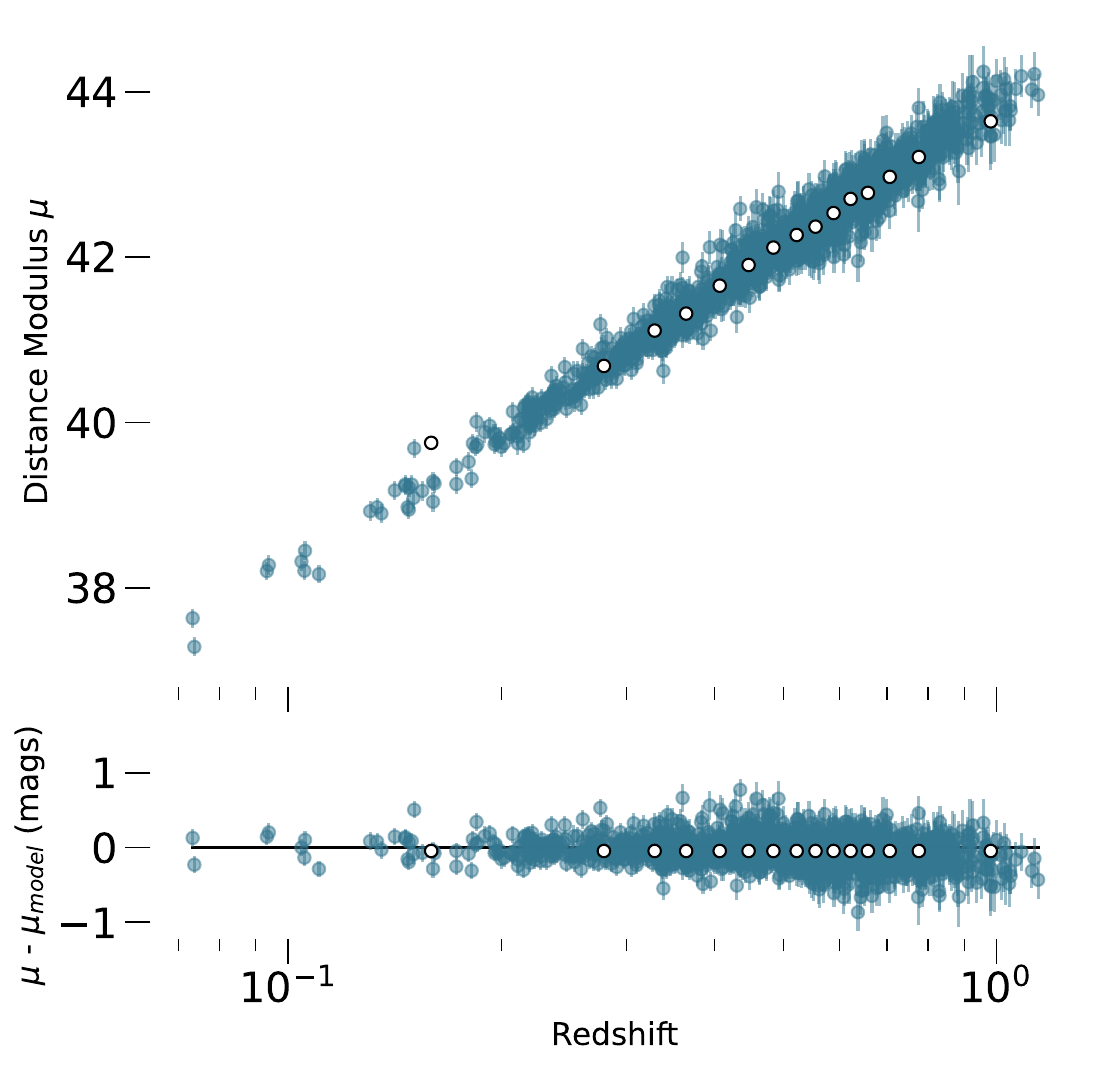}
    \caption{\textbf{Top:} Hubble diagram for DES-SN5YR sample after applying a cut based on photometric classification score. \textbf{Bottom:} As top panel but showing Hubble residuals.}
    \label{fig:DES5YRHubbleDiagram}
\end{figure}

\begin{figure*}
    \centering
    \includegraphics[width = \textwidth]{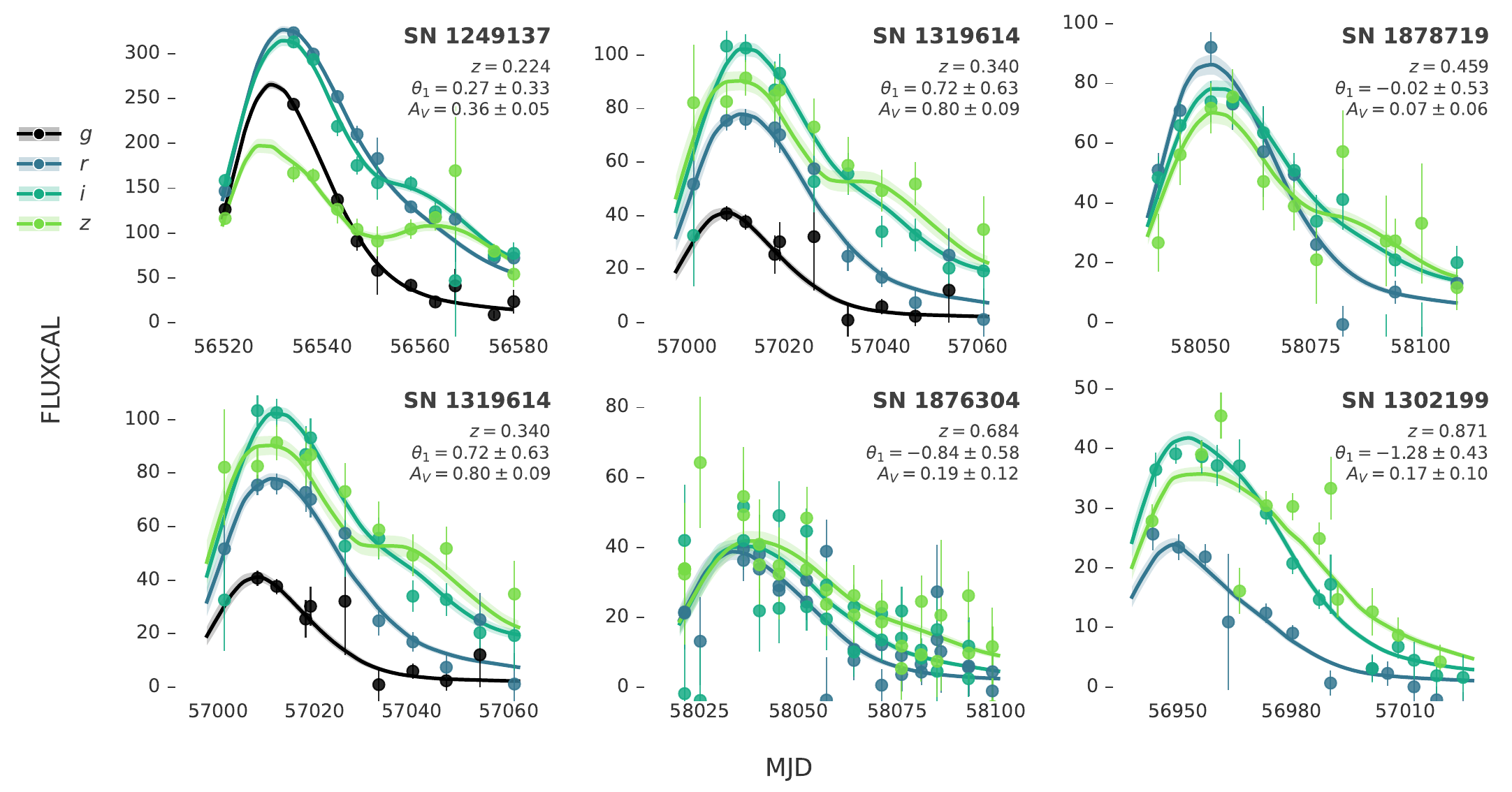}
    \caption{Example light curve fits to some SNe from DES-SN5YR using the G26 BayeSN model, annotated with inferred parameters of each SN.}
    \label{fig:example_lc}
\end{figure*}

While the performance of our updated BayeSN model on the training set is promising, it is also important to assess performance on a data set independent of the one it was trained on. To do so, we apply the G26 model to the DES-SN5YR data from \citet{Sanchez24}. We impose an additional requirement that the photometric classification probability provided in the DES-SN5YR data from the SuperNNova classifier \citep{Moller19, Moller24} is greater than 0.5 ($P_{\rm Ia} \geq 0.5$) to mitigate non-Ia contamination. Fig. \ref{fig:DES5YRHubbleDiagram} shows the Hubble diagram obtained using our model, while Fig. \ref{fig:example_lc} shows some example light curve fits to DES-SN5YR SNe.

We compare the distances inferred using our new BayeSN model to those from SALT3.Dovekie to assess performance. Please note, the published distances for the DES-SN5YR sample include bias corrections; shifts to the inferred distances to correct for biases inherent in the analysis such as Malmquist bias. Work to incorporate BayeSN in the SNANA framework \citep{Kessler16} used to calculate bias corrections is ongoing, but for now to enable an apples-to-apples comparison we compare distances inferred from both models without bias corrections. To mitigate Malmquist bias when comparing the results, we institute a cut of $z < 0.7$, following \citet{Wiseman22, Popovic24b}.

For the G26 BayeSN model, with this sample we find $\sigma_{\rm NMAD} = $ 0.164 mag, compared with $\sigma_{\rm NMAD} = $ 0.185 mag using SALT3.Dovekie\footnote{This is larger than typically reported in DES-SN5YR cosmology papers because of the lack of bias correction.}. Overall, we see a $\sim12$ per cent improvement in scatter relative to the SALT fits with DES-SN5YR. Our conclusions from this comparison are not affected by choosing different redshift cuts for this analysis; these results are seen consistently across the full redshift range. 

\section{Discussion and Conclusions}\label{sec:Conclusions}

In this work, we have provided two key steps towards the BayeSN cosmology effort. First, we introduced the first cross-calibration efforts within the BayeSN framework, allowing for the combination of disparate samples with cosmology-grade precision. At the same time, BayeSN is now able to simultaneously marginalise over filter-specific ZP and wavelength offsets to provide additional constraint on cross-calibration systematics beyond the current approach. Second, we have developed a new model training solution for BayeSN to increase the available training data and extend to higher redshifts.

For this work, we have only included the calibration products from \citetalias{Dovekie} using Gaussian priors. The BayeSN calibration method that we have introduced here, then, only leverages photometry from the SNe Ia to determine the necessary wavelength and ZP shifts, with \citetalias{Dovekie} results acting as an informative prior. A future improvement to the BayeSN calibration would be to simultaneously leverage SN Ia light curves \textit{and} tertiary stars, which would allow for a more accurate accounting of the errors and weights between stars and SNe Ia. Further, the $g-i$ colour of SNe Ia is known to be bluer than stars, and leveraging the entire inferred spectral range from stars and SNe Ia may provide additional information with which to calibrate. However, the broad agreement between our results and the tertiary priors indicates that there is little coherent change between the two spectral ranges, and that mis-calibration arising from the different colour populations is unlikely. Admittedly, there are large changes in a few cases: the CSP$-B$ band shifts by $11.6\sigma$, but this behaviour is not echoed in other $B$ bands in CfA3 or CfA4. We have observed several large ($>5\sigma$) ZP and wavelength shifts, particularly in CfA3K; testing the cosmological impact of these with BayeSN simulations is not yet possible but will be enabled by aforementioned ongoing work to incorporate BayeSN in SNANA.

The new BayeSN training methodology introduced here is an important step towards BayeSN cosmology, enabling training on high-redshift supernovae without introducing a circular dependence on cosmological parameters. This will provide new opportunities for investigating and modelling high-redshift phenomena with BayeSN, paving the road towards the eventual BayeSN cosmology analysis. 

We have stopped before estimating the impact of our photometric uncertainty on cosmological parameters; a proper accounting would require greater mitigation of selection effects than are implemented here \citepalias{Dovekie}. In an ideal setting, cosmological parameters would be inferred directly within the BayeSN model ensuring that these constraints are jointly marginalised over survey cross-calibration; this requires proper treatment of selection effects in the BayeSN model, the focus of ongoing work using SBI. As a more intermediate solution, we can follow a similar approach to \citet{Fragilistic_pub}, \citetalias{Dovekie}, drawing multiple model realisations from the posterior distributions obtained when training the BayeSN model and using each realisation to infer distances. This approach will allow the full model training uncertainty, including cross-calibration uncertainty, to be propagated through to inference of distance and cosmological parameters.

The G26 model will play a key role in incorporating BayeSN within the current multi-step method for cosmological analyses, which involves `correcting' inferred per-SN distances with bias corrections determined using survey simulations. These corrections are applied to account for any biases that arise from the analysis pipeline, including but not limited to those resulting from Malmquist bias and survey selection effects. Within this framework the G26 model will be used to infer SN distances with minimal scatter, while forward simulations will be based on BayeSN-based inferences of the environmental dependence of SNe Ia such as \citet{TM22, Grayling24, GP25}. Most notably, this model has already been used as the basis of an environmental analysis of ZTF in \citet{Ginolin26}. As we enter the LSST era, BayeSN is well-placed to play a key role in the next generation of SN cosmology analyses.

\section*{Acknowledgements}

Supernova and astrostatistics research at Cambridge University is supported in part by the European Union’s Horizon 2020 research and innovation programme under European Research Council Grant Agreement No. 101002652 (BayeSN; PI K. Mandel).
This project has received funding from the European Union’s Horizon Europe research and innovation programme under the Marie Skłodowska-Curie Grant Agreement No. 101205780.

\section*{Data Availability}

All data analysed in this work is publicly available from \url{https://github.com/PantheonPlusSH0ES/DataRelease} for Pantheon+ and \url{https://github.com/des-science/DES-SN5YR} for DES-SN5YR. The G26 BayeSN model used for this analysis is available at \url{https://github.com/bayesn/bayesn}.



\bibliographystyle{mnras}
\bibliography{matt} 

\appendix

\section{Validating Filter Shifts}
\label{appendix:validation}

As part of this work, we validate the performance of our model by injecting artificial shifts into the filter responses of known bands; if we inject a shift of +50 \AA\ for a given band in pre-processing, we would expect to recover a shift consistent with -50~\AA\ when applying our cross-calibration BayeSN model. We choose filters which have shifts consistent with zero in our framework (Foundation $g$-band and PS1 $i$-band) and inject artificial wavelength shifts of +50 \AA\ and -50 \AA\ respectively to test recovery. However, when we do apply this shift we also relax the priors on these individual shifts; for Foundation $g$-band, the prior on the wavelength shift is $\Delta\lambda_T\sim\mathcal{N}(0, 5^2)$ so injecting a 50 \AA\ shift would require the data to be sufficiently constraining to recover a $10\sigma$ shift away from this artificially false prior. As a result, we relax the prior on $\Delta\lambda_T$ specifically for the bands we inject shifts for.

This validation test gives very promising results. For Foundation $g$-band, after injecting a wavelength shift of +50 \AA\ we infer $\Delta\lambda_T = -54.81\pm5.53$ \AA\, and for PS1 $i$-band we infer $\Delta\lambda_T = +42.20\pm6.20$~\AA\ after injecting a shift of -50~\AA. This successfully undoes our artificial shift and recovers an unshifted filter response from the default. Overall, this test gives us confidence that our cross-calibration approach within BayeSN works effectively and gives robust results.


\bsp	
\label{lastpage}
\end{document}